# Experimental Test of the "Isotropic" Approximation for Granular Materials using p=constant Compression

## P. Evesque

Lab MSSMat, UMR 8579 CNRS, Ecole Centrale Paris
92295 CHATENAY-MALABRY, France, e-mail evesque@mssmat.ecp.fr

**Abstract:**

*Experimental data from axially symmetric compression test at constant mean pressure p=($\sigma_1+\sigma_2+\sigma_3$)/3 on Hostun sand from Bouvard experiments are used to study the validity of an "isotropic" modelling as a function of density. The isotropic assumption is found to be quite good for loose samples and/or in the range of large pressure. For smaller mean pressure, anisotropic response is observed at few percents of axial deformation. Relation with anisotropic distribution of local force is made.*

**Pacs # :** 5.40 ; 45.70 ; 62.20 ; 83.70.Fn

______________________________________________________________________

Recent papers [1-3] have proposed a new simple modelling of the behaviour of granular media. It is based on an incremental modelling, assuming an "isotropic" response [4] governed by two plastic parameters, *i.e.* the pseudo Young modulus $1/C_o$ and pseudo Poisson coefficient $\nu$. This modelling assumes then that the incremental response of any compression test with axial symmetry obeys at any stage of the deformation the following equation:

$$\begin{pmatrix} d\boldsymbol{e}_1 \\ d\boldsymbol{e}_2 \\ d\boldsymbol{e}_3 \end{pmatrix} = C_o \begin{pmatrix} 1 & -\boldsymbol{n} & -\boldsymbol{n} \\ -\boldsymbol{n} & 1 & -\boldsymbol{n} \\ -\boldsymbol{n} & -\boldsymbol{n} & 1 \end{pmatrix} \begin{pmatrix} d\boldsymbol{s}_1 \\ d\boldsymbol{s}_2 \\ d\boldsymbol{s}_3 \end{pmatrix} \qquad (1)$$

where $C_o$ plays the part of an inverse pseudo Young modulus and $\nu$ the part of a pseudo Poisson coefficient. $\sigma_i$ states for the effective stress supported by the granular assembly alone; it does not include the pressure $u_w$ supported directly by the liquid invading the pores. It has been proposed in [1-3] that evolution of $C_o$ and $\nu$ has to be determined from any test of triaxial compression with axial symmetry; and it has been found from triaxial compression at constant $\sigma_3=\sigma_2$ that $\nu$ depends only the stress ratio $\sigma_1/\sigma_3$.

Indeed, this "isotropic" assumption may be a crude approximation even when the medium remains homogeneous with axial symmetry, since as the axial deformation proceeds, the distribution of contacts evolves which can induce anisotropy; in turn, this anisotropy modifies the response equation, *i.e.* Eq. (1). In such a case, if one still assumes that the granular medium remains homogeneous and that the symmetry remains axial during the whole deformation, one expects the more general form to be valid:

$$\begin{pmatrix} de_1 \\ de_2 \\ de_3 \end{pmatrix} = -C_o \begin{pmatrix} 1 & -\nu' & -\nu' \\ -\nu & a & -\nu'' \\ -\nu & -\nu'' & a \end{pmatrix} \begin{pmatrix} ds_1 \\ ds_2 \\ ds_3 \end{pmatrix} \quad (2)$$

At this stage, it is worth recalling that $\nu=\nu'$ if the two different compression paths ($\delta\sigma_1 \neq 0$, $\delta\sigma_2=0$, $\delta\sigma_3=0$) and ($\delta\sigma_1=0$, $\delta\sigma_2 \neq 0$, $\delta\sigma_3=0$) pertain to the same incremental linear zone. It is also worth recalling that any axial compression give access to the quantity $\alpha$-$\nu''$ only so that $\alpha$ and $\nu''$ cannot be measured separately with such axi-symmetric triaxial apparatus.

Anyhow, one can then ask whether it is useful to use Equation (2) rather than its simplified "isotropic" version, *i.e.* Eq. (1). Indeed this can be checked directly from experimental data. For instance the validity of the "isotropic" version has been discussed when applied to oedometer compression [3,5]. But this validity can be questioned in a more general way. One way to check its validity is to apply an axially symmetric triaxial compression at constant pressure $p=(\sigma_1+\sigma_2+\sigma_3)/3$, since one expects $\delta v=0$ in the case of an "isotropic" response whatever the applied deviatoric stress $q=\sigma_1-\sigma_2$.

However, to prove the isotropy of the response, it is not sufficient to investigate the variation of the specific volume v versus the deviatoric stress q and to show that v depends very little on q, since applying Eqs. (1) or (2) allows to determine the law of variation of $\varepsilon_v$ vs. q which is expected from the two approaches; this allows to show that both approaches leads to $\delta\varepsilon_v=C_o \cdot f(\nu,\nu',\nu'') \cdot \delta q$ which can be small when either $C_o$, or f, or both are small. So it does not demonstrate that f is small, which is the only requirement for the isotropic response. This is why the procedure proposed in [6] is not fully adapted to test the "isotropy" of the mechanical response.

So, a better way to proceeds is just to use experimental plots of variations of $\varepsilon_v$ as a function of $\varepsilon_1$. Since dp=0 imposes $\delta\sigma_1=(2/3)\delta q=-2\delta\sigma_2=-2\delta\sigma_3$, one gets from Eq. (2):

$$\delta\varepsilon_v = \delta\varepsilon_1 [(1-\alpha)+(\nu'+\nu''-2\nu)]/(1+\nu') \quad (3)$$

When the response is isotropic, one gets $\nu=\nu'=\nu''$ and $\alpha=1$, so Eq. (3) becomes:

$$\delta\varepsilon_v = \delta\varepsilon_1 [O(1-\alpha)+O(\nu'+\nu''-2\nu)]/(1+\nu') \quad (4)$$

where O(x) is a function which tends to 0 as x. Furthermore, an estimate of $\alpha$, $\nu$, $\nu'$ & $\nu''$ are 1, ½, ½ & ½ respectively, since these values are those obtained for the critical state. So, Eq. (4) becomes:

$$\delta\varepsilon_v = (2/3)[O(1-\alpha)+O(\nu'+\nu''-2\nu)] \, \delta\varepsilon_d \quad (5)$$

So Eq. (5) gives the expected variations of $\delta\varepsilon_v$ *vs.* $\delta\varepsilon_1$ for an axially symmetric compression at constant mean pressure p. It indicates that the smaller the slope of $\varepsilon_v$ *vs.* $\varepsilon_1$ the better the "isotropic" approximation.

Fig. 1 reports experimental variations of q and of the density $\rho$ as a function of $\varepsilon_1$ during 3d biaxial compression at constant mean pressure p. These data have been obtained on Hostun sand by D. Bouvard [7], for different initial densities and at different mean pressures. From the $\rho$ data one obtains $\varepsilon_v$ variations according to $\varepsilon_v=-\delta\rho/\rho$.

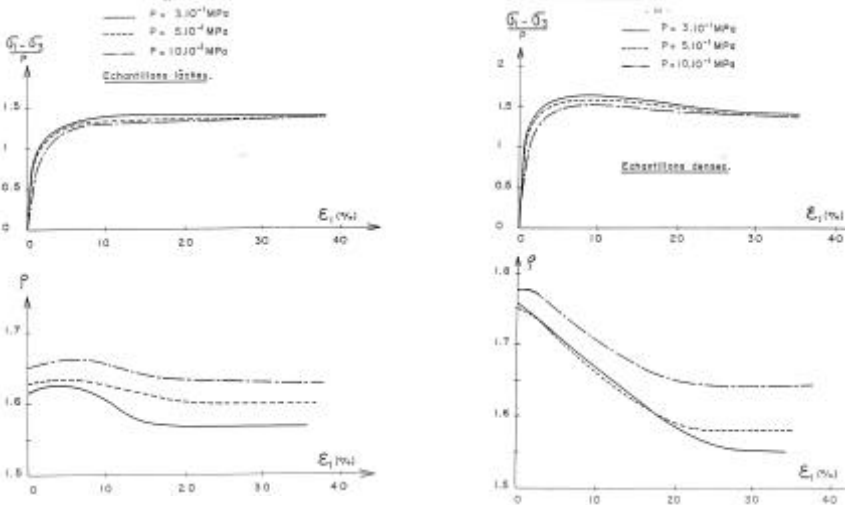

***Figure 1:*** *variations of the deviatoric stress* q/p *(crosses) and of volume deformation* $\varepsilon_v$ *(circles) vs. the axial deformation* $\varepsilon_1$ *during axial compressions at constant mean pressure* p, *for different* p *values and for 2 different initial densities, on Hostun sand. Data from* [7] .

So Fig. 1 shows that :
i) The $\varepsilon_v$ *vs.* $\varepsilon_1$ variations starts with a horizontal tangent at $\varepsilon_1$=0 in general.
ii) The $\varepsilon_v$ *vs.* $\varepsilon_1$ curves remain flat and horizontal in a given range of $\varepsilon_1$ which depends on the initial condition: in the case of the loose sample the $\varepsilon_v$ *vs.* $\varepsilon_1$ curve remains rather flat at all the time for the two larger pressures p, *i.e.* p=0.5 M Pa and p=1 M Pa; it becomes inclined at $\varepsilon_1$>8%. In the case of the dense sample the $\varepsilon_v$ *vs.* $\varepsilon_1$ curve remains flat and horizontal only at small e1 for the two smaller pressures p, *i.e.* p=0.3 M Pa and p=0.5 M Pa, and this behaviour is not observed for the larger pressure (p=1 M Pa)..
iii) Large $\varepsilon_v$ *vs.* $\varepsilon_1$ variations are obtained after a while; the slope becomes important when the q/p is comparable to the critical-state value, *i.e.* q/p=M'=6sin$\varphi$/(3-sin$\varphi$); the slope is about maximum when the q/p *vs.* $\varepsilon_1$ curve is maximum. At this time, the slope $\varepsilon_v/\varepsilon_1 = -\delta\rho/(\rho\ \varepsilon_1)$ becomes quite large and reaches ½ .

iv) After this density variation a steady state is observed at large deformation; it corresponds to the critical state for which $\rho=f(p)$ and $q/p=M'$.
v) At last, one shall remark that no horizontal tangent in the curve $\varepsilon_v$ *vs.* $\varepsilon_1$ is observed for the dense sample under the largest pressure p, *i.e.* p=1 M Pa. It is likely due to the fact that the domain of constant $\rho$ is small and to lack of accuracy of the density measurement.
vi) In the case of the loose and less stressed sample, one observes a small density increase followed by a small density decrease. But these $\rho$ variations fall probably in the error bar on $\rho$.
vii) One observes also the flattest $\varepsilon_v$ *vs.* $\varepsilon_1$ response, which lasts all over the whole range of deformation in the case of the loose sample under intermediate pressure.

So, these data show that "isotropic" approximation is valid in the vicinity, *i.e.* few percent, of $\varepsilon_1=0$. But the size of this vicinity depends on p and r since data show that the denser the sample at a given pressure the sooner the response is "anisotropic" and the smaller the $\varepsilon_1$ at which the slope of $\varepsilon_v$ *vs.* $\varepsilon_1$ becomes important. Furthermore the "anisotropy" of the response occurs always when the q/p ratio approaches the value M' of the critical state.

This point is worth mentioning since it is not obvious from the experimental curves giving $\varepsilon_v$ *vs.* q or $\varepsilon_v$ *vs.* q/p. It is due to the fact that the experimental value of $C_o$ at a given deformation $\varepsilon_1$ and a given pressure p increases strongly when the sample density is increased. as a consequence, The values of **$e_1$ and of $e_v$** which correspond to the same given q/p shall decrease strongly when density increases. In other terms, it means simply that the mean pseudo Young modulus <E> increases strongly when density increases.

A consequence of this is that the procedure proposed in [6] to demonstrate the existence of a bifurcation process during the undrained test, which is based on the analysis of the $\varepsilon_v$ *vs.* q curves obtained from compression tests at constant pressure p, does not assume or does not imply an isotropic response: this demonstration requires only that volume variation remains small when q/p ranges from 0 to M' about, 0<q/p<M'; this is observed experimentally indeed.

Anyhow, points (ii-iii) demonstrate that generation of "anisotropic" response does not require the evolution of contacts in sands at least, since "anisotropic" behaviours appear at small $\varepsilon_1$ already. On the contrary, they mean that the observed "anisotropic" response corresponds mainly to a pure stress-induced "anisotropy", and does not require the generation of strain. This result may then reinforce the results from numerical simulations on granular media obtained by Radjai and coworkers [8] who have observed two force networks, one which remains isotropic and corresponds to an isotropic answer, and the other which is the answer of the granular medium to an anisotropic stress [8] and which generates an anisotropic response. This is important to remark since in my previous papers [2,3,5] I was assuming rather that the development of "anisotropic" answer was requiring the generation of an important axial strain ($\epsilon_1 > 5\%$). On the contrary, the experimental data on sands seem to infirm this hypothesis.

If this was confirmed, it would mean that the mechanical response would only depend in fact on the only two following parameters: the deviatoric stress q and the mean pressure p; in other word it would mean that the response remain isotropic but depend on the q/p ratio, since true anisotropic response require a memory effect which includes the evolution of the contact distribution.

As a conclusion this paper shows from experimental data on axially symmetric compression of Hostun sand at constant pressure that "isotropic" response is valid (i) in the limit of q/p< M' and is no more valid for q/p $\geq \approx 0.9$M'. It is also worth mentioning that $\delta\epsilon_v/\delta\epsilon_1$ can always reach the value ½ which indicates a quite large "anisotropic" response, according to Eq. (5).

At last, it is worth recalling that undrained experiment shows the validity of the "isotropic" approximation till q/p≈M' since the trajectory of this test in the (p,q) plane is a vertical line which turn right on the q=M'p line; it demonstrates also the validity of the notion of characteristic states. But it is demonstrated from compression tests at constant pressure that the "isotropic" approximation is no more valid when q>M'p. These two results seem incompatible, which is quite strange. Does it mean that the q=M'p is quite near a border zone which delimit the change of behaviour for some kind of compression test and not for others. If this interpretation is true, and as the $\epsilon_v$ and q/p *vs.* $\epsilon_1$ variations do not exhibit any abrupt change, it would imply that the delimitation is rather fuzzy or broad: it is better an interface, with a given width, than a surface. Anyhow, more information is needed to conclude surely.

*Acknowledgements:* I want to thank Prof. J. Biarez for helpful discussion and CNES for partial funding.

The electronic arXiv.org version of this paper has been settled during a stay at the Kavli Institute of Theoretical Physics of the University of California at Santa Barbara (KITP-UCSB), in june 2005, supported in part by the National Science Fundation under Grant n° PHY99-07949.

*Poudres & Grains* can be found at :
http://www.mssmat.ecp.fr/rubrique.php3?id_rubrique=402